\documentclass[aps,footinbib,showpacs,twocolumn,superscriptaddress,footinbib]{revtex4}
\usepackage{graphicx,amsmath,color}

\newcommand{\eq}[1]{Eq.~(\ref{#1})}
\newcommand{\fig}[1]{Fig.~\ref{#1}}

\newcommand{\olcite}[1]{Ref.~\onlinecite{#1}}
\newcommand{\olcites}[1]{Refs.~\onlinecite{#1}}

\newcommand{\beq}{\begin{equation}}
\newcommand{\eeq}{\end{equation}}
\newcommand{\bea}{\begin{eqnarray}}
\newcommand{\eea}{\end{eqnarray}}

\newcommand{\bhu}{ {\bf \hat{u}} }
\newcommand{\bhub}{ {\bf \hat{u}}^{\prime} }
\newcommand{\br}{ {\bf r} }
\newcommand{\bhn}{ {\bf \hat{n}} }

\newcommand{\kbt}{k_{\rm B}T}

\begin{document}

\title{Phase diagram of hard colloidal platelets: a theoretical account}
\author{H. H. Wensink}
\affiliation{Department of 
Chemical Engineering, Imperial College London, South Kensington Campus, London SW7 2AZ, United Kingdom}
\author{H. N. W. Lekkerkerker}
\email{h.n.w.lekkerkerker@uu.nl}
\affiliation{Van 't Hoff Laboratory for Physical and Colloid Chemistry, Debye Institute, Utrecht University, Padualaan 8, 3584 CH Utrecht, The Netherlands}

\date{\today} 
\begin{abstract} 
We construct the complete liquid crystal phase diagram of hard plate-like cylinders for 
variable aspect ratio using Onsager's second virial theory with the Parsons-Lee decoupling approximation to 
account for higher-body interactions in the isotropic and nematic fluid phases. The stability of the solid (columnar) 
state at high packing fraction is included by invoking a simple equation of state based on a Lennard-Jones-Devonshire 
(LJD) cell model which has proven to be quantitatively reliable over a large range of packing fractions. By employing an 
asymptotic analysis based on the Gaussian approximation we are able to show that the nematic-columnar transition is 
universal and independent of particle shape. The predicted phase diagram is in qualitative agreement with simulation 
results. \end{abstract}

\pacs{64.70.mf, 82.70.Dd, 05.20.-y}

\maketitle

\section{Introduction}
Many colloidal dispersions, such as natural clays,  and  macromolecular systems consist of oblate or disk-shaped mesogens whose intrinsic ability to form liquid crystalline order gives rise to unique rheological and optical properties. Despite their abundance in nature, the statistical mechanics of fluids containing non-anisometric particles in general (and oblate ones in particular) has received far less attention than that of their spherical counterparts.
 The possibility of a first order disorder-order transition from an isotropic  to a discotic nematic fluid of platelets was first established theoretically by Onsager \cite{Onsager} in the late 1940s. Although originally devised for rod-like particles in solution, his theory also makes qualitative predictions for plate-like particles based on the central idea that orientation-dependent steep repulsive interactions alone are responsible for stabilising nematic order.

The intrinsic difficulty with platelets, as  pointed out by Onsager in his original paper, is that the contribution of third and higher body correlations can no longer be neglected like for thin rod-like species. Consequently,  the original second-virial treatment is expected to give qualitative results at best \cite{Forsyth77}. In a pioneering simulation study, Eppenga and Frenkel \cite{Eppengafrenkel} provided numerical evidence for an isotropic-nematic transition in systems of infinitely thin circular disks and found the transition densities to be much smaller and the first order nature of the transition  to be  much weaker than predicted by the Onsager theory. Owing to the simplicity of the model, the discrepancy can be attributed entirely to the neglect of third and higher body virial terms in the theory.

At high densities, the nematic phase of disk-shaped particles becomes unstable with respect to columnar order, characterised by a planar (2D) hexagonal arrangement of columns each with a liquid internal structure.  Similar to the formation of the nematic phase, the stability of the columnar phase can be explained solely from entropic grounds \cite{frenkellc, Veerman}. Although the system loses configurational entropy because of the partial crystallisation associated with columnar order, this loss is more than offset by a simultaneous increase in translational entropy, i.e. the average available space for each particle increases.

Attempts to improve Onsager's second virial theory have met with variable success (see \olcite{harnauplates} for a recent overview). These  approaches usually involve integral equation or geometric density functional methods whose applicability is often restricted to isotropic fluids  \cite{costahansen,cheung}, models with parallel or restricted orientations \cite{harnaurowan} or particles with vanishing thickness  \cite{esztermann,harnaucosta}. A recent generalisation of the fundamental measure approach towards arbitrarily shaped hard convex bodies provides a potentially promising avenue to address more realistic models for liquid crystal ordering \cite{goosmecke}.  The influence of higher-body correlations can only be assessed numerically via computer simulation \cite{mastersvirial}.  For cut spheres, the virial coefficients have been quantified up to the 8th order both in the fluid isotropic \cite{youvlasov} and nematic state \cite{duncanmasters}. Despite the large number of virial terms, the convergence of the virial expansion of the free energy was found to be insufficient to provide an accurate description of dense nematic and columnar states. Alternatively, Scaled Particle Theory (SPT) can be used to incorporate higher virial terms in an indirect manner. Whilst SPT produces reasonable results for infinitely thin disks \cite{savith}, its extension to finite aspect ratios leads to poor predictions for the isotropic-nematic transition densities \cite{unp}. 

A simpler strategy to account for higher-body particle correlations in the isotropic and nematic fluid state is provided by the so-called Parsons-Lee decoupling approximation \cite{parsons,Lee87,Lee89}. The basic assumption of this approach is that the pair correlation function $g(r)$ of a fluid of hard anisometric bodies, which depends rather intractably on the centre-of-mass distance vector $\Delta \br$  and orientational unit vectors $\bhu$ and $\bhu^{\prime}$, can be mapped onto that of a hard sphere fluid with the same packing fraction via:
\beq
g ( \Delta \br / \sigma_{0} ; \bhu , \bhu^{\prime} ) = g_{\text{HS}} (  \Delta r /  \sigma ( \Delta \hat{\br} ; \bhu, \bhu^{\prime})  ) \label{map}
\eeq
with $\sigma_{0}$ some reference distance (e.g.  particle diameter) and $\sigma ( \Delta \hat{\br} ; \bhu, \bhu^{\prime}) $ the distance of closest approach of a pair of hard anisometric bodies at a given set of orientation unit vectors. In case of hard spheres the distance of closest approach is simply the hard sphere diameter $\sigma_{0}$. \eq{map} provides a natural route of decoupling the translational and orientational degrees of freedom. Starting from the generalised virial equation it is  possible to derive an expression for the excess free energy which is similar to the one from Onsager with the particle density replaced by a {\em rescaled density} involving the hard sphere excess free energy.  Whilst the decoupling approximation is known to work well for short hard spherocylinders \cite{mcgrother}, its merits for plate-like cylinders have not been investigated so far. This we intend to do in the present paper. 

As for the columnar state, the high degree of positional and orientational order can be exploited to devise simple free-volume approaches inspired by  the Lennard-Jones-Devonshire (LJD) cell model \cite{lennardjones,wood,salsburg}. This was first done by Taylor and Hentschke \cite{taylor,hentschke} for the high-density liquid crystal states of parallel cylinders which do not exhibit an isotropic phase.  The approach was further developed and modified in \olcite{wensinkcol} showing that a quantitatively reliable equation of state for the columnar phase can be obtained by accounting for the orientational entropy of the particles, neglected in the original version.
In this paper, we will combine the Onsager-Parsons approach for the isotropic and nematic fluid state with the modified LJD cell theory for the columnar phase to trace the complete phase diagram for freely rotating hard cylinders as a function of thickness-to-diameter ratio. The theoretical predictions will be tested against simulation results for hard cut-spheres. In view of the inherent difficulty of capturing multi-particle correlations in dense plate fluids,  the overall performance of the present theory must be deemed satisfactory. Although quantitative agreement with simulation data is generally lacking, the theory does manage to reproduce the generic features of the phase diagram and provides a simple theoretical underpinning for the relative stability of nematic and columnar order as a function of the plate aspect ratio.

This paper is constructed as follows. Section II and III are devoted to a detailed exposition of the Onsager-Parsons and modified LJD theories, respectively. The phase diagram emerging from the present theory will be presented and discussed in Section IV. Next, algebraic forms of the nematic and columnar free energy are given which allow us to obtain  universal scaling results for the nematic-columnar transition. Finally, some concluding remarks are formulated in Section VI.

\section{Onsager-Parsons theory for the isotropic-nematic transition}

Let us consider a system of hard cylinders  with length $L$ and  diameter $D$ in a macroscopic volume $V$. 
For disk-like cylinders we consider here the {\em aspect ratio} $L/D$  is much smaller than unity. 
The particle concentration is expressed in dimensionless form via $c=ND^3/V$.
Following \olcite{wensinkbidikte}, the Helmholtz free energy within the Onsager-Parsons-Lee approach takes the following form:
\beq
\frac{\beta F}{N} = \ln \tilde{{\mathcal V }} c - 1 + \langle \ln 4 \pi f(\bhu) \rangle 
+ \frac{cG_{P}(\phi)}{2} \frac {  \langle \langle  V_{\text{excl}}(\gamma)  \rangle \rangle }{D^3} \label{free}
\eeq
with $\beta^{-1}=k_{B}T$  the thermal energy ($k_{B}$ represents Boltzmann's constant and $T$ temperature) and $\tilde{{\mathcal V}}= {\mathcal V}/D^{3}$ the dimensionless thermal volume of a platelet including contributions from the rotational momenta.  The brackets $\langle  (\cdot )  \rangle = \int d \bhu f(\bhu) ( \cdot )$, $\langle \langle (\cdot ) \rangle \rangle = \iint d \bhu d \bhub f(\bhu) f(\bhub) ( \cdot )$ denote single and double orientational averages involving some unknown distribution  $f(\bhu)$ of the orientation unit vector $\bhu$ of the plate normal which is normalised according to $\int d \bhu f(\bhu) = 1$. Several entropic contributions can be distinguished in \eq{free}. The first two are  {\em exact} and denote the ideal translational and orientational entropy, respectively. The last term represents the excess translational or packing entropy which accounts for the particle-particle interactions on the approximate level of pair-interactions. The key quantity here is the excluded volume $V_{\text{excl}}$ between two plate-like cylinders at fixed inter-particle angle $\gamma$ with  $\sin \gamma = | \bhu \times \bhub |$. This quantity has been calculated in closed form by Onsager \cite{Onsager} and reads: 
\begin{eqnarray}
\frac{ V_{\text{excl}}( \gamma )}{D^3} &=& \frac{\pi}{2} |\sin \gamma| +  \frac{L}{D}  \left ( \frac{\pi}{2} + 2 E (\sin \gamma) + \frac{\pi}{2} \cos \gamma \right ) \nonumber \\
&&  + 2 \left ( \frac{L}{D} \right )^{2} | \sin \gamma | \label{vexcl}
\end{eqnarray}
with $E(x)$ the complete elliptic integral of the second kind. Although the structure of \eq{free} is similar to the classic Onsager second-virial free energy,  the effect of higher order virial terms are incorporated via  the scaling factor 
\beq
G_{P}=\frac{1-\frac{3}{4}\phi}{(1-\phi)^2}
\eeq
which depends on the total plate volume fraction $\phi = c (\pi/4)L / D$. The rescaled density stems from the Parsons-Lee method \cite{parsons,Lee87,Lee89,wensinkbidikte} which involves a mapping of the plate pair distribution function onto that of a hard sphere system via the virial equation. The free energy can ultimately be linked to the Carnahan-Starling expression for hard spheres, which provides a simple strategy to account for the effect of higher-body particle interactions, albeit in an implicit and approximate manner. As required, $G_{P}$ approaches unity in the limit $\phi \rightarrow 0$ in which case the original second-virial theory is recovered.

Let us now specify the orientational averaging. By definition, all orientations are equally probable in the isotropic (I) phase and $f_{I} = 1/4\pi$. The orientational entropy then vanishes:
\beq
\langle \ln 4 \pi f_{I} \rangle _{I}\equiv 0 
\eeq
If we use the random isotropic averages $\langle \langle \sin \gamma \rangle \rangle_{I}  = \pi/4$, $\langle \langle E(\sin \gamma) \rangle \rangle_{I} = \pi^2 /8 $ and $\langle \langle \cos \gamma \rangle \rangle_{I}  = 1/2$  the excluded volume entropy reduces to:
\beq
\frac{ \langle \langle V_{\text{excl}}(\gamma) \rangle  \rangle _{I} }{D^3} = \frac{\pi^2}{8}  + \left( \frac{3\pi}{4} + \frac{\pi^2}{4} \right ) \frac{L}{D}  +  \frac{\pi}{2} \left ( \frac{L}{D} \right )^{2} 
\eeq
With this, the free energy of the isotropic phase is fully specified.

In the nematic (N) phase, the particles on average point along a common nematic director $\bhn $ and the orientation distribution $f(\bhu \cdot \bhn)$ is no longer a trivial constant. For  a uniaxial nematic phase, $f(\bhu)=f(\theta )$ involving the polar angle  $ 0 \leq \theta \leq \pi$ between the plate normal and the director, with $f$ being a peaked function around $\theta = 0$ and $\theta = \pi$. 

 The equilibrium form follows from the minimum condition of the free energy:
\beq 
\frac{\delta}{\delta f } \left (  \frac{\beta F}{N} - \lambda \int d \bhu f (\bhu ) \right ) = 0  \label{statio}
\eeq
where the Lagrange multiplier $\lambda$ ensures the normalisation of $f$. Applying the condition to \eq{free} leads to a self-consistency equation for $f$:
\beq
f(\bhu ) = \frac{\exp \left [ - cG_{P}(\phi) \int d \bhub  D^{-3} V_{\text{excl}}(| \bhu \times \bhub | ) f(\bhub) \right ] }{\int d \bhu \exp \left [ - cG_{P}(\phi) \int d \bhub  D^{-3} V_{\text{excl}}(| \bhu \times \bhub | ) f(\bhub) \right ]   ]} \label{sc}
\eeq
which needs to be solved numerically for a given particle concentration \cite{herzfeldgrid}. Note that the isotropic distribution $f=1/4\pi$ is a trivial solution of \eq{sc}, irrespective of $c$. At higher densities, nematic solutions of \eq{sc} will appear which give rise to a lower free energy than the isotropic one.  
 The nematic order parameter $S$, defined as: 
\beq
S = \int d \bhu {\mathcal P}_{2} ( \cos \theta)  f(\bhu) \label{s2}
\eeq
[where ${\mathcal P}_{2}(x) = (3x^{2}-1)/2$] is used to distinguish the isotropic state ($S=0$) from the nematic ($0 < S \leq 1$). Once the equilibrium orientational distribution function is known, the pressure $P =  -(\partial F/ \partial V)_{N,T}$ and chemical potential $\mu = (\partial F/ \partial N)_{V,T} $ can be specified to establish phase equilibria between isotropic and nematic states.

\section{LJD cell theory for the columnar phase}

To describe the thermodynamical properties of a columnar phase we use an extended cell theory  as proposed in \olcite{wensinkcol}. 
In this approach, the structure of a columnar phase is envisioned in terms of columns ordered along a perfect lattice in two lateral dimensions with a strictly one-dimensional fluid behaviour of the constituents in the remaining direction along the columns. As for the latter, the configurational integral of a system of {\em parallel} platelets with thicknesses $L$ and  diameter $D$ with their centre-of-mass moving along the plate normal on a line of length $\ell$ is formally given by \cite{tonks}:
\beq
Q_{\text{fluid}} (N , \ell , T ) = \frac{1} {\Lambda ^{N }  N ! }   \left [   \ell -  N L \right ] ^{N} 
\eeq
with $\Lambda$ the thermal de Broglie wavelength. The columns are assumed to be strictly linear and rigid so that fluctuations associated with bending of the columns can be ignored. Next, we allow the platelets to rotate slightly about their centre-of-mass. At high packing fractions, the rotational freedom of each platelet is assumed to be asymptotically small and the configurational integral above may be approximated as follows:
\beq
Q_{\text{fluid}} (N , \ell , T ) \approx \frac{ Q _{\text{or}} } { {\mathcal V}_{1} ^{N}   N ! } \left [   \ell -  N \langle L_{\text{eff}}  \rangle   \right ] ^{N} \label{qrot}
\eeq
where ${\mathcal V_{1}}$  represents the total 1D thermal volume including contributions arising from the 3D rotational momenta of the platelet. Furthermore, $Q_{\text{or}} = \exp [ -N \langle  \ln 4 \pi f \rangle  ] $ is an  orientational partition  integral depending on the orientational probability distribution $f$. In the {\em mean-field} description  implied by  \eq{qrot} there is no coupling between the orientational degrees of freedom of the platelets. The rotational freedom of the platelets is expressed in an {\em effective entropic thickness}, defined as  
\beq
\langle L_{\text{eff}} \rangle  = L \left \{ 1 + \frac{1}{2}\frac{D}{L} \int d(\cos \theta) |\theta| f(\theta)  + \cdots  \right \} \label{leffe}
\eeq
up to leading order in the polar angle $\theta$ which measures the deviation of the plate normal from the direction of the column. 
A prefactor of `$1/2$' in \eq{leffe} has been included to correct in part for the  azimuthal rotational freedom and captures the effect that the excluded length between two platelets at fixed polar angles becomes minimal when the azimuthal orientations are the same. The free energy of the 1D fluid
then follows from $\beta F = -\ln Q$:
\beq
\frac{\beta F_{\text{fluid}}}{N} = \ln \tilde{{\mathcal V}}_{1} \rho  - 1  +  \langle \ln 4 \pi f \rangle  
- \ln \left [ 1 - \rho  \langle \tilde{L}_{\text{eff}} \rangle  \right ] \label{free1d}
\eeq
in terms of  $\tilde{{\mathcal V}}_{1} = {\mathcal V}_{1}/L$, the reduced {\em   linear density}  $\rho = NL/\ell$ and effective thickness $\tilde{L}_{\text{eff}} = L_{\text{eff}}/L$.

Similar to the nematic case in the previous Section, the equilibrium form  $f(\theta)$ is found by a formal minimisation of the free energy under the normalisation constraint. The corresponding stationarity condition is given by \eq{statio}. Since the free energy \eq{free1d} depends only on one-particle orientational averages, the equilibrium ODF can be obtained in closed form and turns out be of a simple exponential form:
\beq
f (\theta) =  \frac{\xi^{2}}{4 \pi} \exp [- \xi |\theta |] \label{expo}
\eeq
with
\beq
\xi = \left ( \frac{3}{2} \frac{D}{L} \right ) \frac{\rho}{1 - \rho }
\eeq
The orientational averages are now easily carried out and the leading order expressions for the orientational entropy and entropic thickness are given by:
\begin{eqnarray}
 \langle \ln 4 \pi f \rangle  &=& 2 \ln \xi  - 2 \nonumber \\
 \langle \tilde{L}_{\text{eff}} \rangle  &=&  1 + \left (\frac{D}{L} \right ) \frac{1}{\xi}
\end{eqnarray}
As a measure for the orientational order along the column, we can define a nematic order parameter $S$ [{\em cf.} \eq{s2}] related to $\xi$ via:
\bea
S  \equiv \left \langle {\cal P}_{2}(\cos \theta)  \right \rangle \sim 1 - \frac{3}{2} \left \langle  \theta^{2} \right \rangle  \sim  1 - \frac{9}{\xi^{2}} \label{xis}
\eea
Let us now turn to the free energy  associated with the positional order along the lateral directions of the columnar liquid crystal. 
A formal way to proceed is to map the system onto an ensemble of $N$ disks ordered into a 2D lattice. Near the close packing density, the configurational integral of the system is provided in good approximation by the LJD cell theory \cite{lennardjones,wood,salsburg,kirkwoodfreevol}.  Within the framework of the cell model, particles are considered to be localised in `cells' centred on the sites of a fully occupied lattice (of some prescribed symmetry). Each particle experiences a potential energy $u_{\text{cell}}^{\text{nn}}({\bf r})$ generated by its nearest neighbours. In the simplest version, the theory presupposes each cell to contain one particle moving {\em independently} from its neighbours.
The $N$-particle canonical partition function can then be factorised as follows:
\begin{eqnarray}
Q_{\text{LJD}}(N) &=& \frac{1} {  \Lambda ^{2N} } \int d{\bf r}^{N} \exp[-\beta U({\bf r}^{N})] \nonumber \\ 
& \approx & \left(  \frac{1} { \Lambda ^{2}} \int d^{2}{\bf r}\exp \left[-\frac{\beta}{2} u_{\text{cell}}^{\text{nn}}({\bf r}) \right] \right)^{N}
\end{eqnarray}
For hard interactions, the second phase space integral is simply the cell {\em free area} available to each particle. If we assume the nearest neighbours to form a perfect hexagonal cage, the free area is  given by $A_{\text{free}}=\sqrt{3}(\Delta_{C}-D)^{2}/2$ with $\Delta_{C}$ the nearest neighbour distance. The configurational integral then becomes \begin{equation}
Q_{\text{LJD}}(N) \approx  \left ( \frac{ A_{\text{free}} }{\Lambda ^2}  \right)^{N} = \left ( \frac{\frac{1}{2}\sqrt{3}\Delta_{C}^{2}} {\Lambda ^2} \right )^{N} \left(  1-\bar{\Delta}_{C}^{-1}   \right)^{2N}
\end{equation}
where the (lateral) spacing $\bar{\Delta}_{C}=\Delta_{C}/D$ is a measure for the translational freedom each particle experiences within the cage. 
The free energy associated with the LJD cell theory is given by:
\beq
\frac{ \beta F _{\text{LJD}} }{N}  = \ln \frac{\Lambda^2}{D^2}  + \ln \frac{2}{\sqrt{3}}  + 2 \ln \left ( \frac{  \bar{\Delta}_{C}^{-1} }{1- \bar{\Delta}_{C}^{-1}}    \right )  \label{fljd}
\eeq
The LJD equation of state associated with \eq{fljd} provides a very accurate description of a 2D solid at densities near close-packing \cite{aldersimul}.
 If we now apply the condition of single-occupancy (i.e. one array of platelets per column) we can use $\bar{\Delta}_{C}$ to relate the  plate volume fraction $\phi = Nv_{0} / V$  (with $v_{0} = (\pi/4) L D^2 $ the particle volume) to the reduced linear density $ \rho $ via:
\begin{equation}
\phi^{\ast} \bar{\Delta}_{C}^{2} = \rho  \label{tienrhophi}
\end{equation}
in terms of the reduced packing fraction $\phi^{\ast} =\phi /\phi_{\text{cp}}$ with $\phi_{\text{cp}}=\pi/2\sqrt{3}\approx 0.907$ the value at close packing. 
The total free energy of the columnar state is now obtained by adding the fluid and LJD contributions:
\begin{eqnarray}
\frac{\beta F_{\text{col}}}{N}  & = &  \ln \tilde{{\mathcal V}}c  - 1 + 2 \ln \left \{  \frac{3}{2} \frac{D}{L} \left ( \frac{\phi^{\ast} \bar{\Delta}_{C}^{2}   }{1 - \phi^{\ast} \bar{\Delta}_{C}^{2} } \right ) \right \} - 2  \nonumber \\
&& - \ln \left (  \frac{ 1 - \phi^{\ast}  \bar{\Delta}_{C}^{2}}{3} \right ) - 2 \ln ( 1 -  \bar{\Delta}_{C}^{-1} )  \label{freecell}
\end{eqnarray}
where the ideal contribution is identical to that of \eq{free}. The final step is to minimise the total free energy with respect to $\bar{\Delta}_{C}$. The stationarity condition $\partial F / \partial \bar{\Delta}_C = 0$ yields a third-order polynomial whose physical solution reads:
\beq
\bar{\Delta}_C = \frac{-3^{1/3}4 \phi^{\ast}  + 2^{1/3} K^{2/3}}{6^{2/3} \phi^{\ast} K ^{1/3}}
\eeq
with 
\beq
K = 27 (\phi ^{\ast}) ^2  + [3 (\phi^{\ast})^3 (32 + 243 \phi^{\ast})]^{1/2} 
\eeq
With this, the free energy for the columnar state is fully specified. Unlike the nematic free energy,  the columnar free energy is entirely algebraic and does not involve any implicit minimisation condition to be solved ({\em cf.} \eq{statio}). The pressure and chemical potential  can be found in the usual way by taking the appropriate derivative of  \eq{freecell}. In Sec. \ref{asymp} we will show that the nematic free energy can also be recast in closed algebraic form using a simple variational form for the ODF, similar to \eq{expo}.

\begin{figure*}
\begin{center}
\begin{picture}(20,0)
\put(1,-5){(a)}
\put(9,-5){(b)}
\end{picture}
\includegraphics[clip=,width = 0.9\columnwidth ]{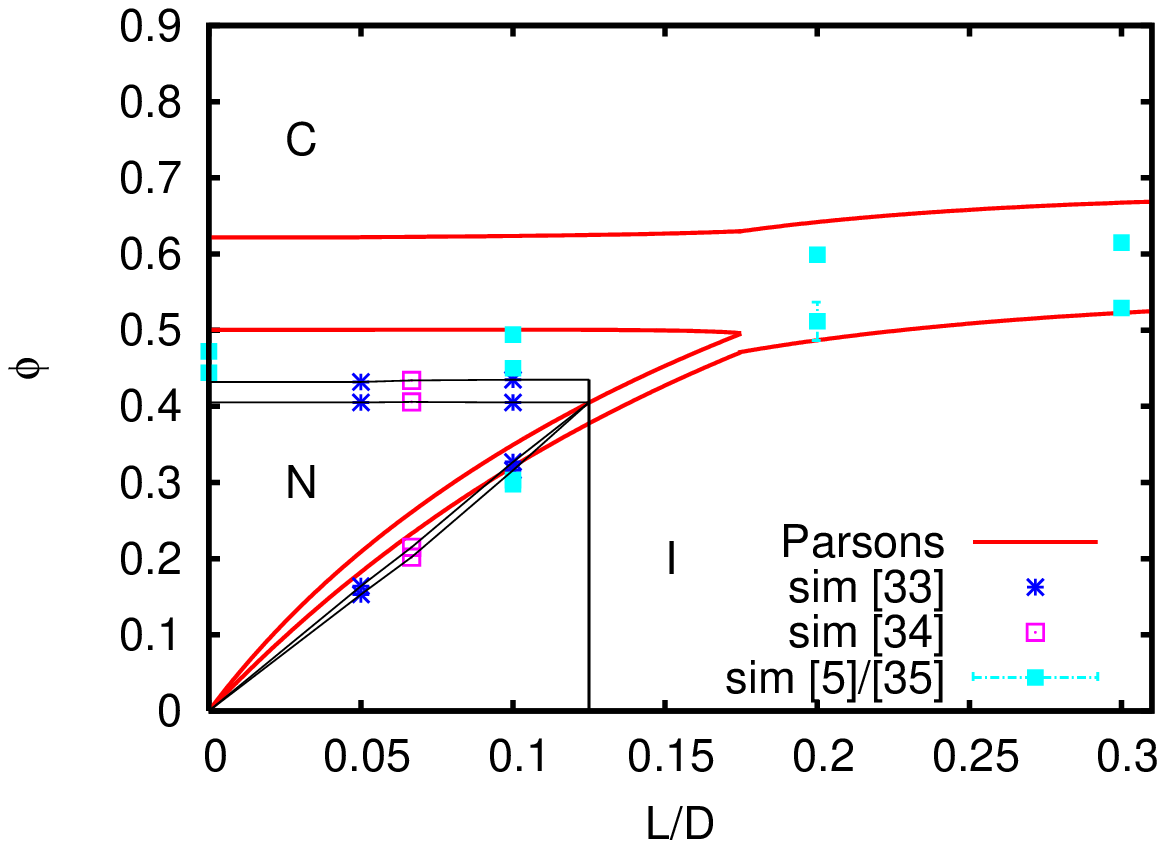}
\includegraphics[clip=,width = 0.9\columnwidth ]{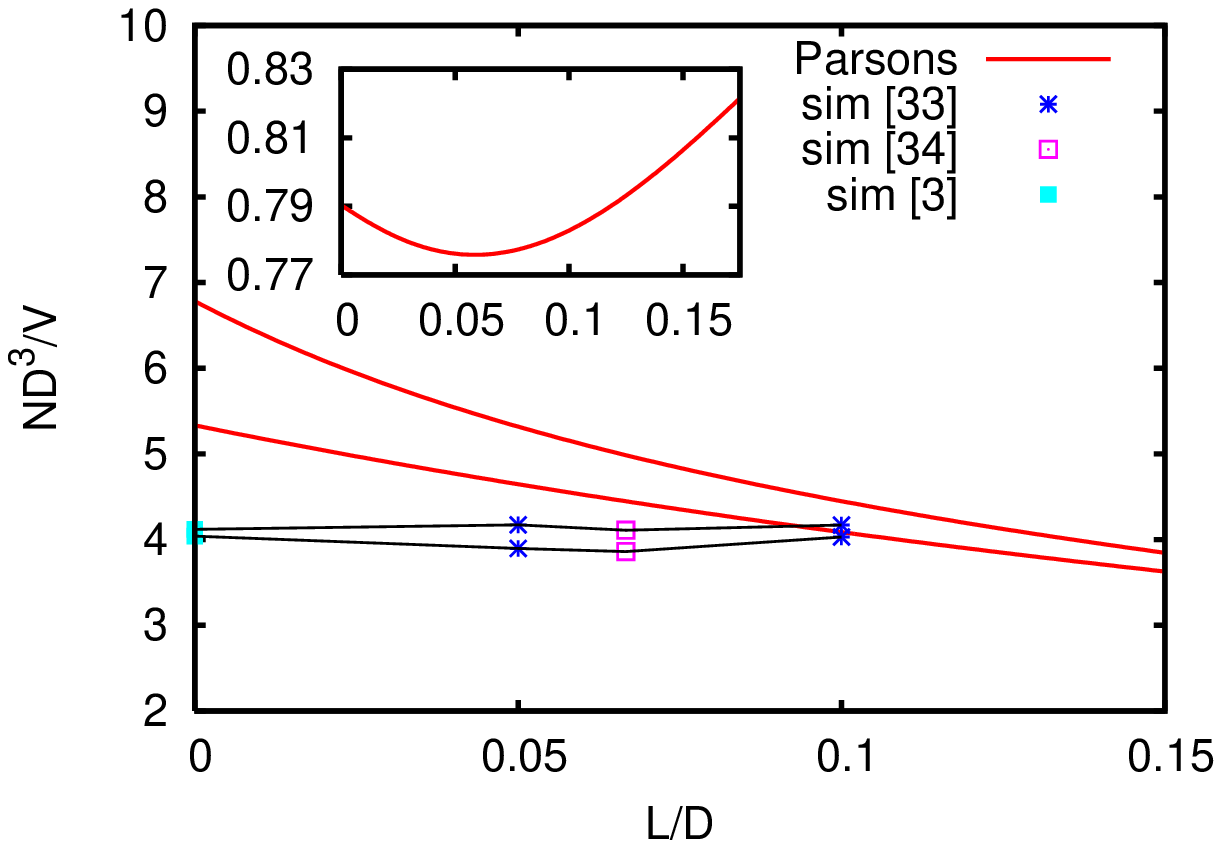}
\caption{\label{mono} (a) Phase diagram for monodisperse colloidal platelets of variable aspect ratio $L/D$ in terms of the  plate packing fraction $\phi =(\pi/4) LD^{2}N/V$. Thin continuous lines serve to guide the simulation data. (b) Dimensionless concentration of the coexisting isotropic and nematic phases as a function of the plate aspect ratio.  Inset: nematic order parameter $S$ of the coexisting nematic phase  plotted versus aspect ratio. }
\end{center}
\end{figure*}

\section{Phase diagram}

\fig{mono} presents an overview of the phase behaviour of a hard cylindrical platelets based on the theoretical approach described above, along with various simulation data for hard cut spheres available in literature. From  \fig{mono}a it is evident that the packing fractions associated with the isotropic-nematic coexistence increase for larger aspect ratio whereas the nematic-columnar transition remains virtually unaffected by the shape of the platelet. This observation is in line with the tentative phase diagram constructed by Veerman and Frenkel \cite{Veerman}. 
The trends can be understood qualitatively by noting that the onset of nematic order occurs if the fraction of {\em excluded volume} $\sim ND^3/V  $ exceeds a certain universal value of about $4$ (as reflected in \fig{mono}b)  whereas columnar order only becomes stable  beyond a critical packing fraction, typically $\phi \simeq 0.4$. Whence:
\begin{eqnarray}
\phi_{\text{IN}} & \simeq  & \pi L/D, \hspace{1cm} L/D \ll 1 \nonumber \\
\phi_{\text{NC}} & \simeq  & 0.4
\end{eqnarray}
which implies the presence of a {\em triple} aspect ratio,  fixed by the intersection of both nematic binodals. 
Although at this particular value  an isotropic-nematic-columnar triphasic coexistence occurs above a certain packing fraction, the system volume occupied by the nematic phase is always infinitesimally small  and the situation thus differs from a regular tri-phasic coexistence occurring in e.g. binary mixtures at  a given thermodynamic state point. Equating both expression we estimate the triple aspect ratio to be $L/D = 0.4\pi \approx 0.126$, which is very close to the value $0.125$ obtained from extrapolating the simulation binodals from \olcites{zhang2,beekschilling}.  Beyond the triple aspect ratio, the platelets are no longer sufficiently anisometric to guarantee a stable nematic phase and direct transitions from the isotropic fluid  to the columnar solid occur. We should note that our theory does not take into account the theoretically disputed cubatic phase, as an intermediate state between the isotropic and columnar phases. The issue of the stability of cubatic order with respect to columnar order is discussed in a recent simulation study by Duncan {\em et al.} \cite{duncanmasters}.   The transition densities from Veerman \cite{Veerman} and Bates \cite{batesthindisks} are  systematically larger than those reported by Zhang \cite{zhang2} and van der Beek \cite{beekschilling} and therefore give rise to a slightly higher estimate of the triple value ($L/D \simeq 0.14$).

\begin{figure*}
\begin{center}
\begin{picture}(20,0)
\put(1,-5){(a)}
\put(9,-5){(b)}
\end{picture}
\includegraphics[clip=,width= 0.9 \columnwidth ]{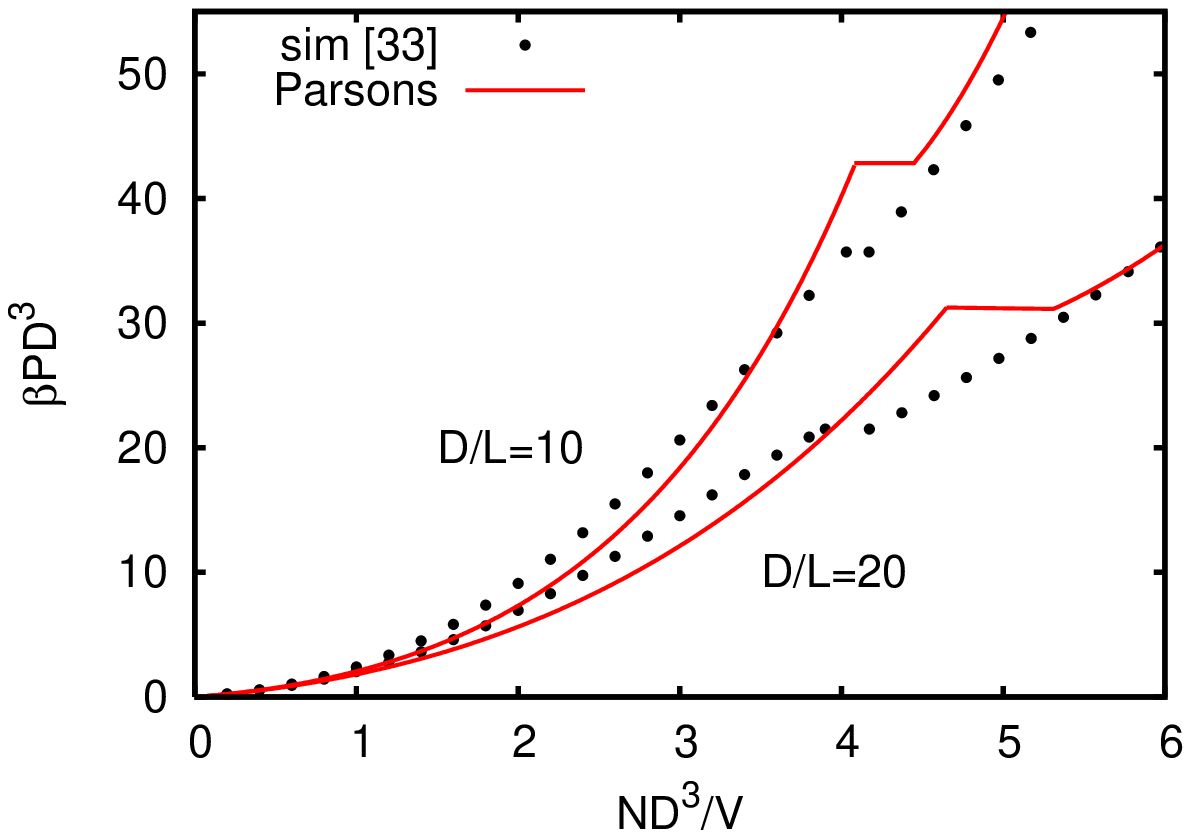}
\includegraphics[clip=,width= 0.9 \columnwidth ]{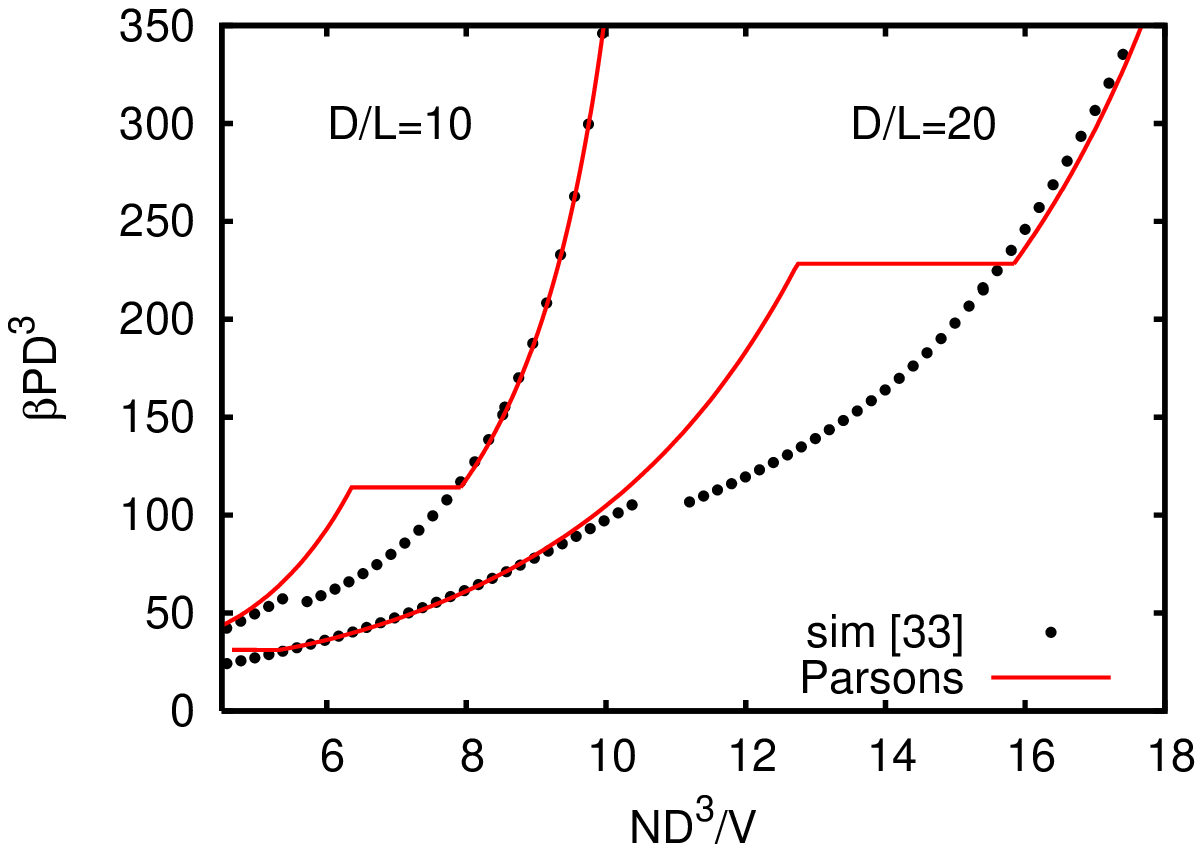}
\caption{\label{druk} Equation of state for colloidal platelets for two different inverse plate aspect ratios $D/L$, plotted in terms of the reduced pressure $PD^{3}/\kbt$ versus dimensionless concentration $ND^{3}/V$. (a) Isotropic-nematic density region. (b) Nematic-columnar region.}
\end{center}
\end{figure*}

The theoretical value $L/D =0.175$ deviates considerably from the ones predicted from simulations, mainly because the predicted packing fractions of the coexisting nematic and columnar phases are too large. The equations of state presented in \fig{druk}  demonstrate that the main source of error must be the chemical potential of either the nematic or columnar phase nematic branch, rather than the pressure. For $L/D=0.05$, the predicted pressure in the nematic and columnar states are fairly close to the simulation results with discrepancies  less than a few percent in both branches. 

For larger aspect ratios ($L/D > 0.1$) the agreement between theory and simulation is quite satisfactory, despite an increased shape difference between the cylinder and the cut sphere. For $L/D = 0.2$, the occurrence of cubatic order has been reported in simulation \cite{Veerman} which is not taken into account in the present model. The isotropic binodal point in \fig{mono}a at this value is taken  to be the mean value between the onset of cubatic order and the transition to the columnar state, with the error bar indicating the boundaries of cubatic order. For the thickest species $L/D=0.3$, the coexisting high density phase was found to be a solid rather than a columnar. Similarly, for smaller aspect ratios a continuous columnar-solid transition line could be located beyond the nematic-columnar moving toward higher packing fraction upon decreasing $L/D$. In our simple cell-fluid model there is, however, no distinction between the columnar and solid states due to the absence of a freezing transition within the strictly 1D line fluid representing the structure along the column direction.

The predictive power of the Onsager-Parsons theory for platelets is perhaps better highlighted in \fig{mono}b, where the isotropic-nematic binodals are plotted in terms of the reduced number concentration $c=ND^{3}/V$. The agreement is reasonable for large aspect ratio but rather poor for thin platelets. In the limit of infinitely thin disks ($L/D \rightarrow 0$) the coexistence concentrations are identical to those obtained from Onsager's second-virial theory {\em viz.}  $c_{I} = 3.29 (16/\pi^{2})$ and $c_{N}=4.191(16/\pi^{2})$ \cite{Lekkerkerker84}. This is easily understood from the fact that the packing fraction of infinitely thin disks at a given finite number concentration is zero. Consequently,  $G_{P}(\phi)$ reduces to unity and the Parsons decoupling approximation involving the hard sphere excess free energy becomes ineffective. A similar reduction to the $B_{2}$ level takes place for infinitely thin rods ($L/D \rightarrow \infty$). However, contrary to rods, the effect of the {\em third virial coefficient} $B_{3}$ is finite for disks with vanishing thickness. In general, for isotropic systems of hard cylinders we have: 
\bea
B_{3} / B_{2}^{2} &=& 0,     \hspace{1cm} ( L/D \rightarrow \infty) \nonumber \\
B_{3} / B_{2}^{2}  &=& 0.444,  \hspace{1cm} ( D/L \rightarrow \infty) 
\eea
where the latter value is taken from \olcite{Eppengafrenkel}. Likewise, higher order virial contributions will also be non-zero. Simulation studies of the virial terms up to $B_{7}$ \cite{Eppengafrenkel,youvlasov} reveal that higher order virial terms involve alternating positive and negative contributions of comparable magnitude, indicating just how complicated virial expansions are for dense fluids of platelets. The virial terms generated by the Onsager-Parsons free energy can be obtained from the virial expression for the excess free energy $\beta F^{\text{ex}}/N = \sum_{n \geq 2} B_{n} \rho^{n-1} / (n-1)$. Applying this to \eq{free} gives:
\beq
\frac{B_{n}}{ B_{2}^{n-1}} = \frac{(n+2)(n-1)}{4(n-2)!} \left ( \frac{\pi}{8} \frac{L}{D} \frac{ \langle \langle  V_{\text{excl}}(\gamma) \rangle \rangle }{D^{3}} \right )^{n-2}\hspace{0.1cm} (n \geq 2) 
\eeq
It is clear that all contributions beyond $B_{2}$ are zero for $L/D=0$ thus leading back to the original Onsager result. For $L/D=0.1$ the reduced third, fourth and fifth virial coefficients in the isotropic phase are 0.170, 0.010 and 3.67 $\cdot 10^{-4}$. Comparing these with the numerically exact values 0.508, 0.111 and -0.10  for cut spheres \cite{youvlasov} shows that higher-order correlations are systematically under-weighted by the Parsons method.

\section{Asymptotic results for the N-C  transition}
\label{asymp}

A simple rationale for the apparent independence of the NC transition with respect to  particle shape can be obtained by comparing the free energy of the two states and exploiting the fact that the nematic order at densities close to the transition is very strong. In that case the average excluded volume \eq{vexcl} between the particles in the nematic phase can be approximated by retaining the leading order contribution for small inter-rod angles $\gamma$:
\beq
\langle \langle \tilde{V}_{\text{excl}}( \gamma ) \rangle  \rangle  \sim 2 \pi  \frac{L}{D} + \left (  \frac{\pi}{2} + \frac{2L^2}{D^2} \right ) \langle \langle \gamma \rangle \rangle
\eeq
The orientational averages indicated by the brackets can be estimated using a Gaussian Ansatz for the ODF \cite{OdijkLekkerkerker}:
\beq
\label{gaussint}
\left  \langle (\cdot) \right \rangle  \sim  \int_{-\pi/2}^{\pi/2} d \theta | \sin \theta | \int_{0}^{2\pi} d \varphi f_{G}(\theta) (\cdot)
\eeq
in terms of the following one-parameter Gaussian variational function:
\beq
f_{G}(\theta) = {\mathcal N} \exp \left [ - \frac{ \alpha }{2} \theta ^{2}  \right ], \hspace{1cm} \left ( -\frac{\pi}{2} \leq \theta \leq \frac{\pi}{2} \right )
\eeq
where the normalisation factor ${\mathcal N}$ follows from $\langle 1 \rangle  =1$. The variational parameter $\alpha$ is required to be much larger than unity so that $f_{G}$ is sharply peaked around $\theta=0$. In that case, the integration over the polar angle $\theta$ in \eq{gaussint} can be safely extended to $\pm \infty$ and $\sin \theta \approx \theta$. In the asymptotic limit, the normalization constant is given by ${\mathcal N}=\alpha/4\pi$ and the double orientational average over the angle $\gamma$ in the nematic phase is found to be  \cite{OdijkLekkerkerker}:
\beq
\left  \langle \left \langle \gamma \right \rangle \right \rangle \sim \left ( \frac{\pi}{\alpha} \right )^{1/2}
\eeq
to leading order in $\alpha $. Similarly, the orientational entropy can be approximated by:
\beq
\left \langle \ln 4 \pi f_{G} \right \rangle \sim \ln \alpha - 1
\eeq
The nematic order parameter [{\em cf.} \eq{xis}] follows from $ S \sim 1-3/\alpha $.
These algebraic results allow the minimisation of the free energy with respect to $\alpha$ to be carried out analytically and leads to a closed expression for the nematic free energy.

Using the asymptotic expressions above, and introducing the volume fraction as a density variable the following algebraic form for the Onsager-Parsons nematic free energy can be produced:
\begin{eqnarray}
\frac{\beta F_{\text{nem}}}{N} & \sim & (\ln \phi - 1) +\left \{ 2 \ln \left ( \frac{D}{L} \frac{\pi^{1/2}}{2} \phi G_{P}(\phi) \right ) -1 \right \}  \nonumber \\
&& + 2 + 4 \phi G_{P}(\phi) 
\end{eqnarray}
which is a sum of the ideal, orientational and excess parts, respectively. Similarly, one may derive for the columnar free energy:
\begin{eqnarray}
\frac{\beta F_{\text{col}}}{N} & \sim  & (\ln \phi  - 1 ) +\left \{ 2 \ln \left ( \frac{3}{2} \frac{D}{L} \frac{\rho}{1 - \rho} \right ) - 2 \right \} \nonumber \\
&&  - \ln \frac{(1-\rho)}{3}  - 2 \ln (1-\bar{\Delta}_{C}^{-1}) 
\end{eqnarray}
with $\rho = \phi^{\ast} \bar{\Delta}_C^{2}$ and combines the ideal, orientational, 1D fluid and cell contributions, respectively. The only explicit shape dependency is the contribution $2 \ln D/L$ in the orientational part which is identical  in  both expressions and therefore does {\em not} affect the  NC coexistence properties. Solving the coexistence conditions gives the universal coexistence values $\phi_{N} = 0.500 $ and $\phi_{C} = 0.621 $ and pressure $\beta P D^{3} = 11.37 (D/L)$. Furthermore, the normalised lateral columnar spacing is $\bar{\Delta}_{C} = 1.075$ and the equilibrium variational parameters $\alpha = 1.232 (D/L)^2$ pertaining to the nematic order in the nematic phase  and columnar phases are given by  $\alpha = 1.276 (D/L)^2$ and  $\xi = 5.830 (D/L)$, respectively.

\section{Conclusions}

We have combined the Onsager-Parsons theory  with a simple LJD cell model to address the phase behaviour of hard cylindrical platelets with variable aspect ratio. 
 The theoretical framework provides a simple, yet qualitative underpinning for the competitive stability of the isotropic, nematic and columnar states, observed in Monte Carlo computer simulations. Upon increasing the aspect ratio, the window of stability for the nematic phase decreases systematically  up to a critical value, identified as a triple equilibrium. Beyond this value the anisometry of the plates is too small to warrant a stable nematic phase and direct transitions from an isotropic fluid to a columnar solid occur. It would be intriguing to verify whether the stability of cubatic order, suggested by the simulations, can be captured with a similar free volume concept.  Rather than forming a closed packed assembly of columns, the cubatic phase  must be envisioned in terms of interacting finite-sized stacks with random orientations. This will be explored in a future study.

\acknowledgments
We are grateful to George Jackson and Jeroen van Duijneveldt for fruitful discussions. HHW acknowledges the Ramsay Memorial Fellowship Trust for financial support.

\bibliography{rik}
\bibliographystyle{apsrev}

\end{document}